\documentclass{aipproc}
\usepackage[latin1]{inputenc}
\usepackage{aipxfm}

\layoutstyle{8x11single}

\begin{document}

\title{Extraction of the Compton Form Factor $\mathcal{H}$ \\ from DVCS Measurements in the Quark Sector}

\author{H. MOUTARDE, on behalf of the CLAS group at Saclay}{address = {CEA, Centre de Saclay, IRFU/Service de Physique Nucléaire, F-91191 Gif-sur-Yvette, France}, email = {herve.moutarde@cea.fr}}

\keywords{Generalized Parton Distributions, Compton Form Factors, Deeply Virtual Compton Scattering.}
\classification{13.60.Fz, 14.20.Dh}

\begin{abstract}
Working at twist 2 accuracy and assuming the dominance of the Generalized Parton Distribution $H$ we study  the helicity-dependent and independent cross sections measured in Hall A, the beam spin asymmetries measured in Hall B at Jefferson Laboratory and beam charge, beam spin and target spin asymmetries measured by Hermes. We extract the real and imaginary parts of the Compton Form Factor $\mathcal{H}$, the latter being obtained with a 20--50~\% uncertainty. We pay extra attention to the estimation of systematic errors on the extraction of $\mathcal{H}$. We discuss our results and compare to other extractions as well as to the popular VGG model.
\end{abstract}

\maketitle


\section*{Introduction}

The Deeply Virtual Compton Scattering (DVCS) process  \cite{Ji:1996ek, Diehl:1997bu} was early recognized as the cleanest way to access Generalised Parton Distributions (GPD) and has been so far a very active field of research. Theoretical developments are reviewed in Refs.~\cite{Goeke:2001tz, Diehl:2003ny, Belitsky:2005qn, Boffi:2007yc} and relevant experimental results are published in Refs.~\cite{Airapetian:2001yk, Adloff:2001cn, Stepanyan:2001sm, Chekanov:2003ya, Chen:2006na, MunozCamacho:2006hx, Aaron:2007cz, Girod:2007jq, Airapetian:2008jga, Airapetian:2008rj, Airapetian:2010mb, Airapetian:2009bm, Airapetian:2009bi}. 

What we would like is to eventually produce a experimental 3d picture of the nucleon thanks to the knowledge of GPDs. Even if getting a precise experimental knowledge of GPDs is a long-term program, we can already perform the first fits to get a flavour of the actual sensitivity of observables to GPDs and elaborate robust and efficient extracting methods. We address these issues by studying recent JLab data, namely beam spin asymmetries  \cite{Girod:2007jq} (BSA) and helicity-dependent and independent cross sections  \cite{MunozCamacho:2006hx}$_{\vphantom{1}}$. These data offer the interesting features of a large kinematic coverage and a fine kinematic binning. We also consider Hermes data~: beam charge, beam spin and target spin asymmetries  \cite{Airapetian:2008jga, Zeiler:2008zy}.

The first section of this paper describes the evaluation of the $ep \rightarrow ep\gamma$ cross sections and outlines our hypothesis. In the second part we explain our fitting strategy. In the third section we discuss the extracted values of $\mathcal{H}$ and compare them to other similar studies.


\section{Preliminary analysis}

\subsection{DVCS at leading twist}

Four GPDs $H$, $E$, $\tilde{H}$ and $\tilde{E}$ appear at twist 2, but the cross sections depend on the Compton Form Factors (CFF) $\mathcal{H}$, $\mathcal{E}$, $\tilde{\mathcal{H}}$ and $\tilde{\mathcal{E}}$. The convention of Ref.~\cite{Belitsky:2001ns} is used to define the CFFs. The complex integration kernel yields a real and an imaginary part to a CFF. For example the CFF $\mathcal{H}$ satisfies~:
\begin{eqnarray}
	Re \mathcal{H} & = & \mathcal{P} \int_{-1}^{+1} dx \, H(x,\xi,t) \left( \frac{1}{\xi-x} - \frac{1}{\xi+x} \right) \label{EqRealPartHE} \\
	Im \mathcal{H} & = & \pi \Big( H(\xi,\xi,t) - H(-\xi,\xi,t) \Big) \label{EqImaginaryPartHE} 
\end{eqnarray}
where $\xi = x_B (1+t/(2Q^2)) / (2-x_B+ x_B t/Q^2))$ is the generalised Bjorken variable  \cite{Belitsky:2001ns, Guichon:2008}$_{\vphantom{1}}$, $x_B$ the standard Bjorken variable, $Q^2$ the virtuality of the initial photon and $t$ the square momentum transfer between initial and final protons. $\mathcal{P}$ denotes the principal value of the integral.

The use of dispersion relations relating real and imaginary parts of a CFF had been discussed by Anikin and Teryaev  \cite{Teryaev:2005uj, Anikin:2007tx, Anikin:2007yh}$_{\vphantom{1}}$, Diehl and Yvanov  \cite{Diehl:2007jb} and Goldstein and Liuti  \cite{Goldstein:2009ks, Goldstein:2009rb, Goldstein:2010ce}$_{\vphantom{1}}$. Dispersions relations are used in the model-dependent global fit of Kumeri\v{c}ki and Müller  \cite{Kumericki2009uq}$_{\vphantom{1}}$. However the unknown subtraction (the D-term  \cite{Polyakov:1999gs}) and the limited kinematic range of available data make this constraint rather weak in model-independent fitting. In this work we try to minimize the model-dependence of the extraction. Then we consider real and imaginary parts of CFFs as independent except when we use an explicit parametrisation of GPDs to compute CFFs.

\subsection{$ep \rightarrow ep\gamma$ observables at twist 2}

The harmonic analysis of $ep \rightarrow ep\gamma$ cross-sections has so far relied on the 2002 work of Belitsky, Müller and Kirchner  \cite{Belitsky:2001ns} (BMK). In this formalism, the interference between the Bethe-Heitler (BH) and DVCS processes was treated with a leading order approximation of the BH part. This assumption was removed by Belitsky and Müller in the case of a spinless target  \cite{Belitsky:2008bz} and for a (longitudinally polarised or unpolarised) spin 1/2 target  \cite{Belitsky:2010jw} and by Guichon and Vanderhaegen in the case of a proton target  \cite{Guichon:2008}$_{\vphantom{1}}$. In all the following, we will use the expressions from the latter. At last, the twist 2 approximation is motivated by the claim of early $Q^2$-scaling observed in Hall A  \cite{MunozCamacho:2006hx}$_{\vphantom{1}}$. Moreover the $A^{LU}_{DVCS}$ asymmetry measured by Hermes  \cite{Zeiler:2008zy} are small if not zero while they are supposed to vanish exactly at twist 2.

\subsection{About current extractions methods}

The methods to extract GPDs from measurements fall today into 4 groups~:
\begin{description}
\item[local fits] The real and imaginary parts of CFFs are the free parameters. A given kinematic bin ($x_B, t, Q^2)$ is considered independantly of the others. The model dependence is low but some assumptions on the kinematics may be necessary to avoid underconstrained problems  \cite{Guidal:2008ie, Guidal:2009aa, Guidal:2010ig, Guidal:2010de}. It does not allow any extrapolations and in particular not those necessary to obtain spatial information in the transverse plane  \cite{Burkardt:2000za} or the quark orbital angular momentum in the nucleon  \cite{Ji:1996ek}.
\item[global fits] CFFs or GPDs have model-dependent expressions, the free parameters of which are fitted. See for example  \cite{Kumericki2009uq} for recent fit of DVCS data on unpolarized targets or  \cite{Goloskokov:2005sd, Goloskokov:2007nt, Goloskokov:2009ia} for recent studies of Deeply Virtual Meson Production data.
\item[hybrid fits] This is a combination of the two previous methods, used in  \cite{Moutarde:2009fg} and detailed in the present text.
\item[neural networks fits] Work is in progress but preliminary results were recently presented  \cite{Kumericki2010fr}.
\end{description}

\subsection{$H$-dominance hypothesis}

When we consider experiments on unpolarised proton targets we can neglect $E$, $\tilde{H}$ and $\tilde{E}$ ($H$-dominance). The contribution of these GPDs to helicity-dependent cross-sections is indeed kinematically suppressed  \cite{Moutarde:2009fg}$_{\vphantom{1}}$ and we can check (for instance thanks to the VGG model~  \cite{Guichon:1998xv, Vanderhaeghen:1998uc, Vanderhaeghen:1999xj, Guidal:2004nd}) that $Im \mathcal{E}$ and $Im \tilde{\mathcal{H}}$ are usually smaller than $Im \mathcal{H}$. At last for small $t$ and $\xi$ we expect $\tilde{H}/H$ to be close to $\Delta q/q$ \textit{i.e. } $\frac{1}{4}$. 

Thus assuming $H$-dominance we may hope to extract information on $H$ from BSAs or helicity-dependent cross sections with a systematic error of 20~\% to 50~\%\footnote{A direct test of the $H$-dominance assumption with the VGG model gives an upper bound of 25~\%. This is comparable to the typical statistical uncertainty on BSAs.}, this approximation being better at small $t$. The advantage of this approach is the dramatic decrease of the number of degrees of freedom involved in fits. Guidal indeed showed  \cite{Guidal:2008ie} in the same kinematic region that it is not possible to extract sensible information about the real and imaginary parts of all CFFs by direct fits of helicity-dependent and independent cross sections.


\section{Fitting strategies}

The possibility to study GPDs in DVCS rests on factorisation theorems~  \cite{Collins:1998be} which require a small value of $|t|/Q^2$. In the following, we restrict ourselves to kinematic configurations for which $|t|/Q^2 < 1/2$ and justify this choice \textit{a posteriori}. As stated in the first section, we perform two kinds of fits~:
\begin{description}
	\item[Local fits] $Re \mathcal{H}$ and $Im \mathcal{H}$ are the free parameters of the fits.
	\item[Global fit] We use a "dual model"-like parametrization~; it consists basically in a simultaneous expansion in Gegenbauer and Legendre polynomials. See Ref.~\cite{Moutarde:2009fg} for details on the fitting Ansatz.
\end{description}
We estimate the systematic errors associated to our $H$-dominance hypothesis by first fitting data setting the subdominant GPDs to 0, then fitting the same data setting the subdominant GPDs to their VGG value, and computing the difference. We estimate the systematic uncertainty related to the global fitting Ansatz by varying the number of polynomials involved in the expansion.

A careful fitting strategy is elaborated to handle the high number of free parameters involved in the fits and make sure that the numeric implementation is under control. We adopt an iterative (5 steps) fitting procedure. See Ref.~\cite{Moutarde:2009fg} for full details.


\section{Results}

\subsection{Extraction of $Im \mathcal{H}$ and $Re \mathcal{H}$}

\begin{figure}
		\begin{tabular}{cc}
			\includegraphics[width=0.45\textwidth]{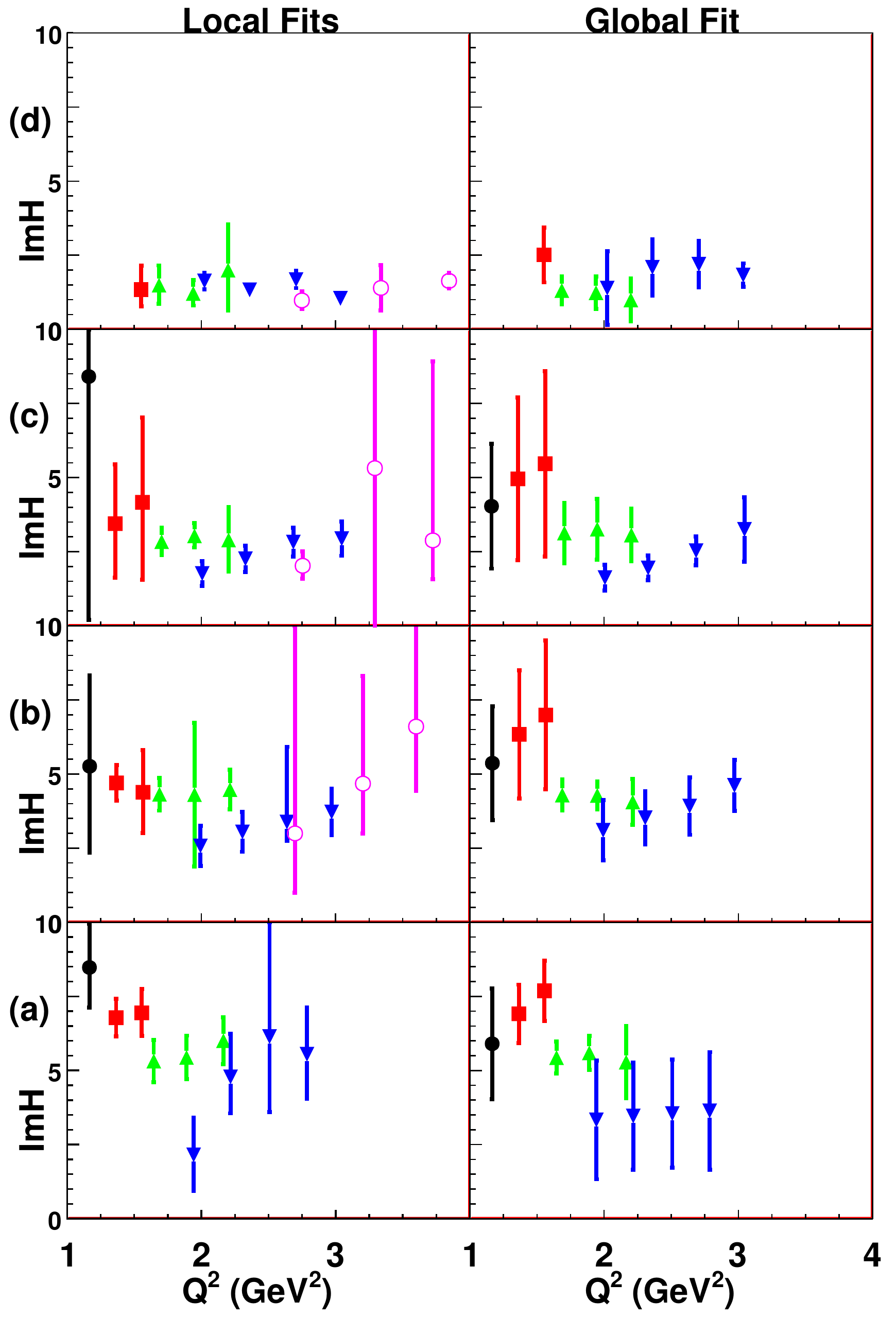}
			&
			\includegraphics[width=0.45\textwidth]{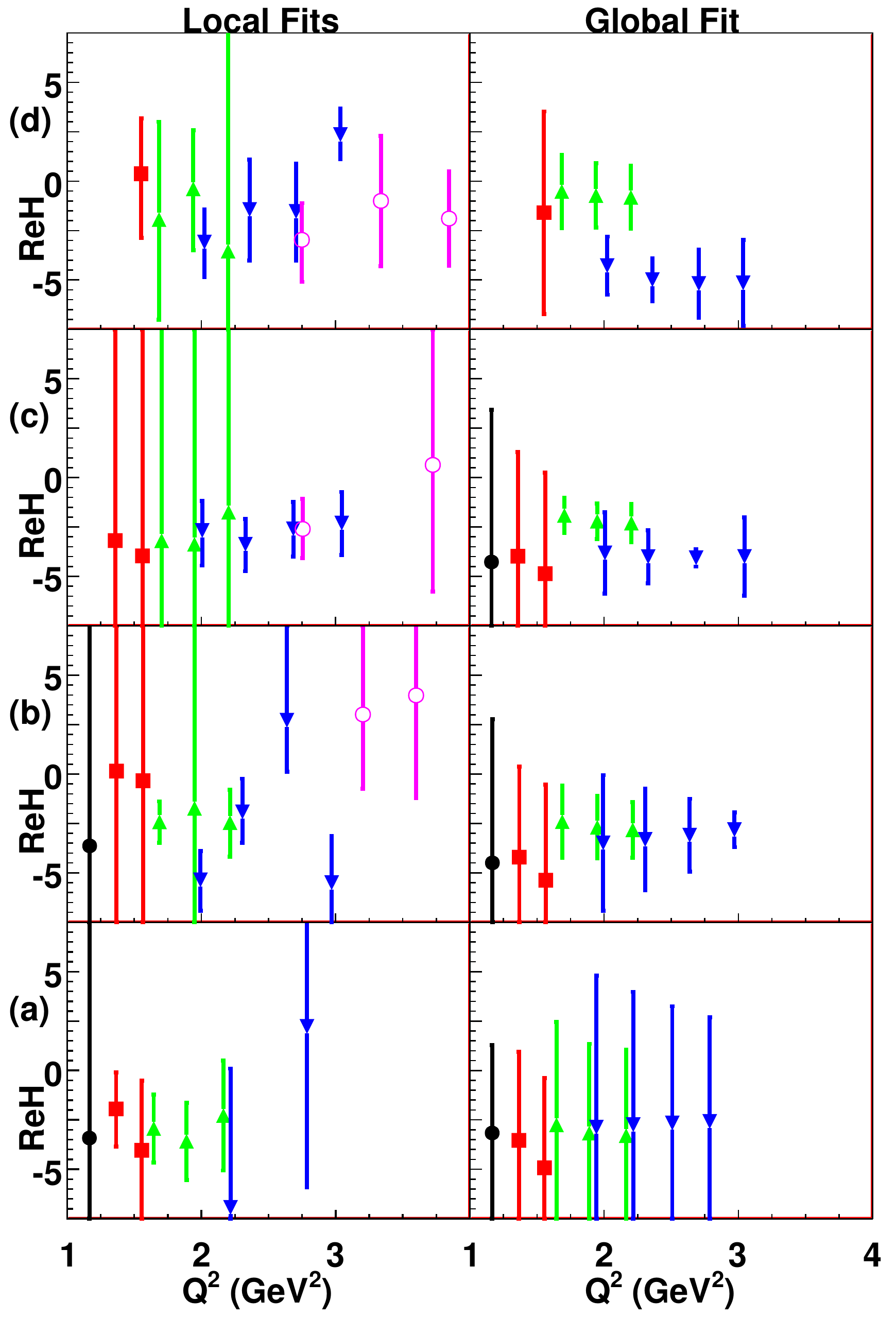}
		\end{tabular}
		\caption{\label{FigScalingImHHallB}$Q^2$-behaviour ($1 < Q^2 < 4~\textrm{GeV}^2$) of the extracted values of $Im \mathcal{H}$ (left) and $Re \mathcal{H}$ (right) of local fits and global fit on Hall B kinematics~: $0.09 < -t < 0.2~\textrm{GeV}^2$ (a), $0.2 < -t < 0.4~\textrm{GeV}^2$ (b), $0.4 < -t < 0.6~\textrm{GeV}^2$ (c), and $0.6 < -t < 1.~\textrm{GeV}^2$ (d). The error bars include both statistics and systematics. $Im \mathcal{H}$ ranges between 0 and 10 and $Re \mathcal{H}$ ranges between -7.5 and +7.5. The black full circles correspond to $x_B$=0.125, red squares to $x_B$=0.175, green up triangles to $x_B$=0.250, blue down triangles to $x_B$=0.360 and magenta open circles to $x_B$=0.491.}
\end{figure}

The Fig.~\ref{FigScalingImHHallB} displays our results for $Im \mathcal{H}$ and $Re \mathcal{H}$. Both local fits and global fit give results with comparable accuracy for $Im \mathcal{H}$, but as expected the results of the global fits are smoother. This is especially true concerning $Re \mathcal{H}$~: in this case the local fits suffer from large fluctuations of $Re \mathcal{H}$ with values which fall outside the plot range. However, we could not reliably extract values of the CFFs for the larger values of $\xi$ with the global fit.

The results for local and global fits are almost always compatible, which is a strong consistency check. Both rely on the assumptions of twist 2 accuracy and of $H$-dominance. On one hand, local fits suffer from numerical fluctuations (the 2-parameter local fits are not constrained enough on some bins) but are almost model-independent. On the other hand, global fits are smoother, but suffer from oscillations. That both methods give the same results indicates that fluctuations and oscillations are reasonably controlled in the bins for which results are displayed. Since, in both cases, the total error bars have the same size, we conclude that our estimation of systematic uncertainties on the global fit Ansatz is realistic even if not perfect.

All fits keep data satisfying $|t|/Q^2<1/2$. For local fits, changing the maximal value of $|t|/Q^2$ amounts to dropping points. For global fits, the whole results may be changed,  but the good agreement between the results of both types of fits, and the slow $Q^2$-evolution of the extracted CFFs, indicate that this restricted kinematic region is suitable for an analysis in the GPD framework.

Our results are explicitly given in Ref.~\cite{Moutarde:2009fg}. Error bars are dominated by systematic effects. Typically we obtain a relative accuracy of 20 to 50~\% on $Im \mathcal{H}$, which is quite satisfactory under the assumption of $H$-dominance and given the statistical accuracy of JLab data. On the contrary, $Re \mathcal{H}$ is still largely undetermined, and is never extracted with a precision better than 50~\%.

\subsection{Discussion}

\begin{figure}
		\includegraphics[width=0.9\textwidth]{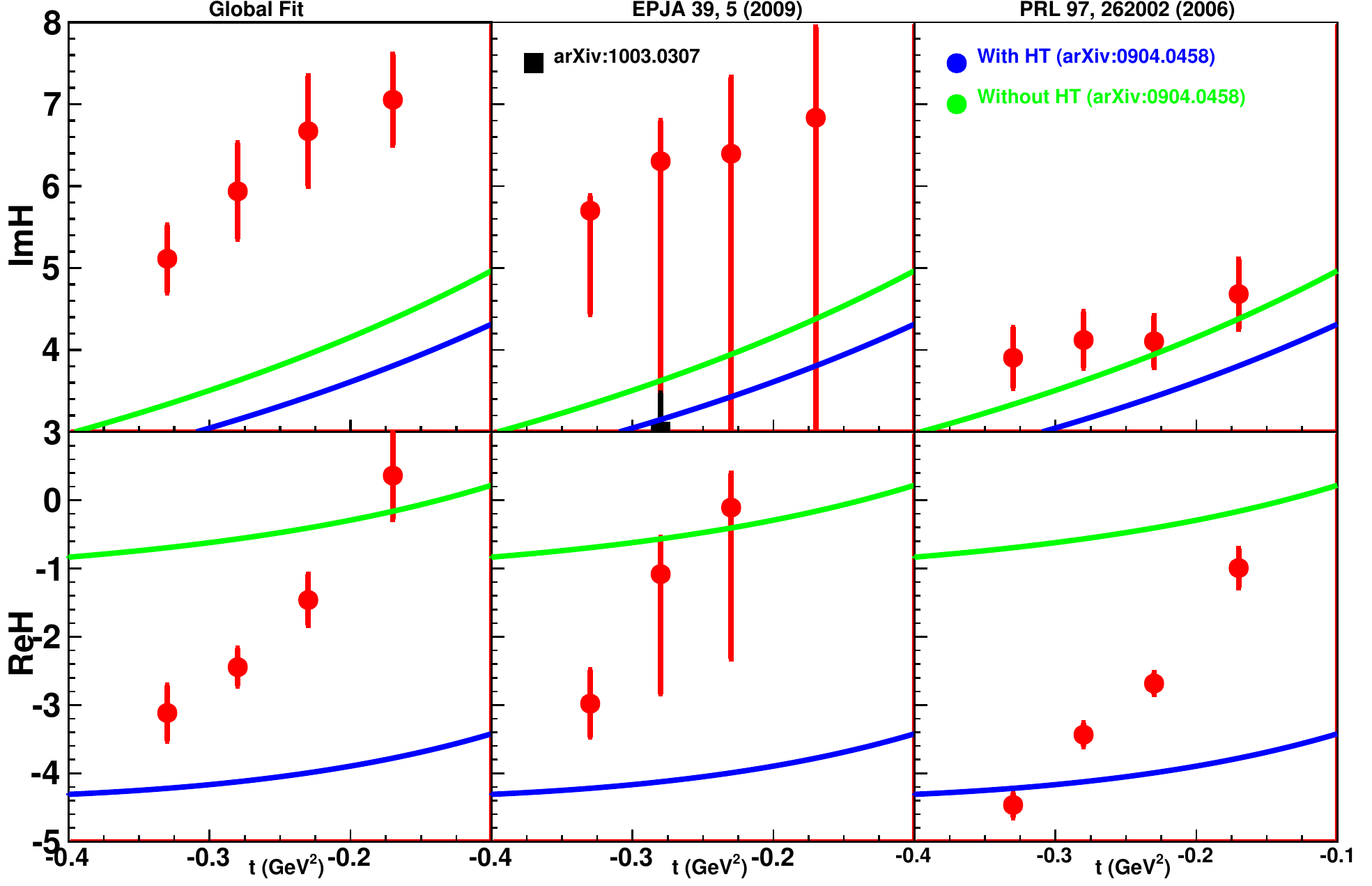}
	\caption{\label{FigCompReHImHHallA}$Im \mathcal{H}$ (up) and $Re \mathcal{H}$ (down) vs $t$ ($0.1 < -t < 0.4~\textrm{GeV}^2$) on Hall A kinematics ($x_B$=0.36 and $Q^2$ = 2.3~GeV$^{2}$). $Im \mathcal{H}$ ranges between 3 and 8, and $Re \mathcal{H}$ between -5 and 1. We compare our results (left column) to those of Guidal  \cite{Guidal:2008ie} (middle column) and Mu{\~n}oz-Camacho \textit{et al.}  \cite{MunozCamacho:2006hx} (right column). In the latter column $\mathcal{H} + \frac{x_B}{2-x_B} \left( 1 + \frac{F_2}{F_1} \tilde{\mathcal{H}} \right) - \frac{t}{4 M^2} \frac{F_2}{F_1} \mathcal{E}$ is plotted, and not $\mathcal{H}$. The error bars include both statistical and systematic uncertainties. The green (resp. blue) curve is the result of the model-dependent global fit of Kumeri\v{c}ki and Müller  \cite{Kumericki2009uq} with (resp. without) $\tilde{\mathcal{H}}$. The full black square is the result of Ref.~\cite{Guidal:2010ig}.}
\end{figure}

The Fig.~\ref{FigCompReHImHHallA} compares our results on Hall A kinematics to twist 2 model-independent extractions  \cite{Guidal:2008ie, Guidal:2010ig} of Guidal and two extractions with the BMK formalism  \cite{MunozCamacho:2006hx, Kumericki2009uq}$_{\vphantom{1}}$. Firstly, the use of the GV expressions creates important deviations to the extraction of Ref.~\cite{MunozCamacho:2006hx}$_{\vphantom{1}}$. Since the extracted combinations of GPDs are not the same, we will not make the argument more quantitative.  Another discrepancy occur when comparing to the results of the global fit of the GPD $H$ of Kumeri\v{c}ki and Müller  \cite{Kumericki2009uq}$_{\vphantom{1}}$. Secondly, we obtained results in good agreement with Ref.~\cite{Guidal:2008ie}$_{\vphantom{1}}$. Note however the surprisingly big shift between the results of Ref.~\cite{Guidal:2008ie} and Ref.~\cite{Guidal:2010ig} induced by the inclusion of target spin asymmetries. These points need to be elucidated but we can already mention the following points~:
\begin{itemize}
\item Large contributions of $\tilde{H}$ and $\tilde{E}$ are necessary in order to fit JLab Hall A data with the parametrization of Kumeri\v{c}ki and Müller. Other DVCS data on unpolarized targets can be included in this global fit under the $H$-dominance assumption.
\item The shift between the two extractions of Guidal is attributed to a $\tilde{H}$ contribution.
\item Using VGG and the MIT Bag GPD model  \cite{Ji:1997gm} we obtained a refined estimate of the systematic uncertainty caused by $H$-dominance. We obtained larger error bars for the extracted CFF on Hall A kinematics, and error bars compatible with previous estimates on Hall B kinematics.
\end{itemize}
These facts indicate that JLab Hall A data can certainly not be understood under the restricted set of assumptions~: twist 2 + $H$-dominance. Confirming an explanation involving GPDs other than $H$ will require the inclusion of observables more sensitive to these GPDs. Moreover it was already stressed in  Ref.~\cite{Moutarde:2009fg} that the twist 2 assumption need to be checked further. Interestingly the three different methods described in Refs.~\cite{Kumericki2009uq, Guidal:2008ie, Guidal:2010ig, Moutarde:2009fg} give results in good agreement on Hall B data. The reason of the distinction (discrepancy for Hall A data, agreement for Hall B data) need to be elucidated to make progress on GPD extraction in the valence region. Let us remind that JLab Hall A and Hall B data are compatible.

\section*{Conclusions}

Working at leading twist, and assuming $H$-dominance, we extracted $Im \mathcal{H}$ and $Re \mathcal{H}$ with 20--50~\% accuracy on $Im \mathcal{H}$. The good agreement between the results of local and global fits is a strong consistency check. Moreover values of $Im \mathcal{H}$ on Hall B kinematics extracted by three different groups with three different methods are in good agreement. The comparison of local and global fits and the weak $Q^2$-dependency of the results also validates \textit{a posteriori} the restriction to kinematic configurations with $|t|/Q^2<1/2$. Note however that the interpretation of JLab Hall A measurements is difficult and probably requires to go beyond the common $H$-dominance approximation.

Our results are dominated by systematic uncertainties. The systematic uncertainties related to $H$-dominance will decrease in the future using additional BSA measurements~\cite{Kubarovsky:2010} on unpolarised and longitudinally polarised proton target which will put stronger constraints on the global fits. These future measurements will also help clarifying the interpretation of Hall A helicity-independent cross sections.


\section*{Acknowledgments}

The author would like to thank the CLAS group at Saclay, P.~Guichon, M.~Guidal, K.~Kumeri\v{c}ki, D.~Müller and K.~Passek-Kumeri\v{c}ki for many fruitful and stimulating discussions. The author also thank the organizers of the 12th International Conference on Meson-Nucleon Physics and the Structure of the Nucleon held in College of William and Mary (May 31-June 4, 2010), Williamsburg, Virginia. This work was supported in part by the Commissariat à l'Energie Atomique and the GDR n° 3034 Physique du Nucleon.


\end{document}